\documentclass{pasj01}
\def\gridline#1{\vskip6pt\hbox to\hsize{#1}\vskip6pt}

\usepackage{url}
\usepackage{ulem}

\Received{$\langle$reception date$\rangle$}
\Accepted{$\langle$acception date$\rangle$}
\Published{$\langle$publication date$\rangle$}

\begin{document}

\title{Numerical Study of Stellar Core Collapse and Neutrino Emission Using the Nuclear Equation of State Obtained by the Variational Method}
\author{Ken'ichiro Nakazato}%
\altaffiltext{}{Faculty of Arts and Science, Kyushu University, 744 Motooka, Nishi-ku, Fukuoka 819-0395, Japan}
\email{nakazato@artsci.kyushu-u.ac.jp}
\author{Kohsuke Sumiyoshi}%
\altaffiltext{}{National Institute of Technology, Numazu College, Ooka 3600, Numazu, Shizuoka 410-8501, Japan}
\author{Hajime Togashi}%
\altaffiltext{}{Department of Physics, Tohoku University, Sendai 980-8578, Japan}


\maketitle

\begin{abstract}
Core-collapse simulations of massive stars are performed using the equation of state (EOS) based on the microscopic variational calculation with realistic nuclear forces. The progenitor models with the initial masses of $15M_\odot$, $9.6M_\odot$, and $30M_\odot$ are adopted as examples of the ordinary core-collapse supernova with a shock stall, the low-mass supernova with a successful explosion, and the black hole formation, respectively. Moreover, the neutrinos emitted from the stellar collapse are assessed. Then, the variational EOS is confirmed to work well in all cases. The EOS dependences of the dynamics, thermal structure, and neutrino emission of the stellar collapse are also investigated.
\end{abstract}

\section{Introduction}
The property of nuclear matter is one of the key components in resolving the collapse of massive stars, which is their ultimate fate~\citep[for example]{2005ApJ...629..922S,2013ApJ...764...99S,2014EPJA...50...46F,2019ApJ...887..110S,Schneider:2019shi}. When the central density exceeds the nuclear saturation density owing to the collapse, the nuclear force becomes repulsive and halts the collapse. Then, a shock wave is launched. If the nuclear matter has a more easily compressible equation of state (EOS), the bounce occurs deeper inside the core. The composition of nuclear matter is relevant for neutrino interactions. Since core-collapse simulations require sufficiently large range of density, temperature, and proton fraction, there are not many EOS models currently available in the numerical simulations. While \citet{1991NuPhA.535..331L} presented an EOS model based on a Skyrme-type interaction in the form of programming routines, several EOS models based on mean field approaches have recently been implemented in table form \citep[for example]{2010NuPhA.837..210H,2011ApJS..197...20S,2013ApJ...774...17S,2015PASJ...67...13I}. We refer the reader to \citet{2017RvMP...89a5007O} for the diversity of EOS models.

Recently, \citet{2017NuPhA.961...78T} have constructed a new EOS for nuclear matter, denoted as the Togashi EOS hereafter, on the basis of variational many-body theory. The Togashi EOS stands out from the other EOS models mentioned above in terms of realistic nuclear potentials, as described in \S\ref{sec:togashieos}. It is tabulated to be applicable not only for core-collapse simulations but also for various astrophysical simulations such as the thermal evolution of isolated neutron stars for about $10^6$~yr~\citep{2019PTEP.2019k3E01D} and binary neutron star mergers~\citep{2019ApJ...884...40M}. In particular, the Togashi EOS has been applied to the cooling simulation of a proto-neutron star, which is a nascent hot neutron star born in a core-collapse supernova, by \citet{2018PhRvC..97c5804N}. The heavy nuclei in the inhomogeneous phase at low densities have larger fractions and larger mass numbers in the Togashi EOS model than in the other EOS models. Thereby, the coherent scattering off the heavy nuclei is enhanced and neutrinos are trapped near the proto-neutron star surface. As a result, the Togashi EOS model has a higher neutrino average energy and a longer duration of neutrino emission in the late time over $O(10)$~s.

In this study, we apply the Togashi EOS to the core collapse of three progenitor models: the models with initial masses of $9.6M_\odot$ and $15M_\odot$ are for core-collapse supernovae and the model with $30M_\odot$ is for black hole formation. The numerical simulations of stellar core collapse and neutrino emission are performed using the neutrino-radiation hydrodynamics code described in \S\ref{sec:sumiyoshicode} under spherical symmetry. It is known that the explosion is unlikely in spherical symmetry because of the stalling of the shock wave, as demonstrated with the $15M_\odot$ model in \S\ref{sec:15m}. Nevertheless, the collapse of the $9.6M_\odot$ model results in a successful explosion due to the low mass accretion rate shown in \S\ref{sec:z9.6}. We also investigate the black hole formation due to a failed supernova with the $30M_\odot$ model in \S\ref{sec:bhmodel}. For each model, we report the impacts of EOS, comparing the results of the Togashi EOS model with those of other EOS models, which would also be informative for other astrophysical applications. We provide the concluding remarks in \S\ref{sec:concl}.

\section{Setup}
\subsection{Equation of state}
\label{sec:togashieos}
For numerical simulations of core-collapse supernovae, 
we use a thermodynamically consistent EOS based on the microscopic variational calculation with realistic nuclear forces\footnote{See also a series of papers~\citep{2007NuPhA.791..232K, 2009PThPh.122..673K, 2013NuPhA.902...53T, 2014PTEP.2014b3D05T, 2017NuPhA.961...78T} for further details.}.
As in the well-known EOS constructed by \citet{1998PhRvC..58.1804A}, hereafter, APR EOS,
the EOS of uniform nuclear matter is calculated starting from  
the nuclear Hamiltonian $H$ including realistic two- and three-body nuclear forces:
\begin{equation}
H= -{\textstyle\sum\limits_{i=1}^N} \frac{\hbar^2}{2m}\nabla_i^2 + {\textstyle\sum\limits_{i<j}^N}V_{ij} + {\textstyle\sum\limits_{i<j<k}^N}V_{ijk}. 
\end{equation}
Here, $m$ is the rest mass of a neutron, $V_{ij}$ is the Argonne v18 two-nucleon potential \citep{1995PhRvC..51...38W}, 
and $V_{ijk}$ is the Urbana IX three-nucleon potential \citep{1995PhRvL..74.4396P}. 
The expectation value of the one- and two-body components of $H$ is evaluated in the two-body cluster approximation with the Jastrow wave function, 
and the obtained two-body energy is minimized with respect to state-dependent correlation functions included in the Jastrow wave function.
The contribution of the three-body energy is also considered referring to the expectation value of the three-body component of $H$ with the Fermi-gas wave function.  
The obtained total energies per nucleon for pure neutron matter and symmetric nuclear matter are in good agreement with 
the results of the APR EOS with sophisticated Fermi hypernetted chain variational calculations \citep{1998PhRvC..58.1804A}.

The uniform matter EOS at zero temperature can be generalized to finite temperature by using the variational method proposed by Schmidt and Pandharipande \citep{1979PhLB...87...11S, 2007PhRvC..75c5802M}. 
The mixing of alpha particles is also considered in this step, where alpha particles are assumed to be a noninteracting Boltzmann gas with a fixed volume. 
For nonuniform nuclear matter, the Thomas--Fermi calculation is performed as in the case of the Shen EOS \citep{2011ApJS..197...20S}.
In this calculation, a single species of spherical heavy nuclei is assumed to form a body-centered cubic lattice surrounded by a gas comprising nucleons and alpha particles. Finally, the thermodynamically favorable state is chosen by comparing the free energy of nonuniform matter with that of uniform matter. 
The obtained EOS is publicly available\footnote{\url{http://www.np.phys.waseda.ac.jp/EOS/}} 
as a table of thermodynamic quantities over a baryon density range from $10^5$ to $10^{16}$~g~cm$^{-3}$, 
a proton fraction range from 0 to 0.65, and a temperature range from 0 to $10^{2.6}$~MeV.

The resultant EOS satisfies the properties of nuclear matter obtained through the terrestrial experiments and astrophysical observations.
At the saturation density $n_0=0.16$~fm$^{-3}$, the saturation energy $E_0$ of symmetric nuclear matter is $-16.09$~MeV, the incompressibility $K$ is 245~MeV, the symmetry energy $E_{\mathrm{sym}}$ is 29.1~MeV, and the symmetry energy slope parameter $L$ is 38.7~MeV,
which satisfy the constraints from terrestrial nuclei \citep[for example]{2018PrPNP.101...55G,2019EPJA...55..117L}\footnote{The values of $E_{\mathrm{sym}}$ and $L$ shown in this paper are slightly different from those previously given in the paper~\citep{ 2017NuPhA.961...78T}. While $E_{\mathrm{sym}}$ in the previous paper is defined as the difference between the energy per nucleon for symmetric nuclear matter and for pure neutron matter, $E_{\mathrm{sym}}$ in this paper is calculated as the second order differential of the energy per nucleon for symmetric nuclear matter by the original definition.}.
The maximum mass of neutron stars calculated using the Togashi EOS is $2.21 M_{\odot}$, which is compatible with the observed masses of heavy neutron stars: J1614$-$2230 with $1.928\pm0.017 M_{\odot}$~\citep{2016ApJ...832..167F}\footnote{The original mass measurement was $1.97\pm0.04 M_{\odot}$ in \citet{2010Natur.467.1081D}.}, J0348+0432 with $2.01\pm0.04 M_{\odot}$~\citep{2013Sci...340..448A}, and J0740+6620 with $2.14^{+0.10}_{-0.09} M_{\odot}$~\citep{2020NatAs...4...72C}. The radius of a $1.4 M_{\odot}$ neutron star is 11.6~km, which is consistent with the LIGO/Virgo inference of $11.9 \pm 1.4$~km~\citep{2018PhRvL.121p1101A} from the binary neutron star merger event GW170817 \citep{2017PhRvL.119p1101A}. Incidentally, some models have recently been proposed, in which the Togashi EOS was adopted. For the high-density region, the Togashi EOS was extended so as to treat the additional hyperon mixing \citep{2016PhRvC..93c5808T} and the transition from hadronic to quark matter \citep{2019ApJ...885...42B,2020PhRvD.102b3031X}. For the low-density region, the nuclear statistical equilibrium was also considered \citep{2017JPhG...44i4001F}.

\subsection{Neutrino radiation hydrodynamics}
\label{sec:sumiyoshicode}
We perform numerical simulations of general relativistic neutrino-radiation hydrodynamics under spherical symmetry. The numerical code directly solves the Boltzmann equation for neutrino distributions by the discrete ordinate method and handles hydrodynamics at the same time in an implicit manner. The numerical simulations start from the initial model of massive stars and follow the time evolution of the gravitational collapse, the core bounce, and the propagation of the shock wave. Depending on the progenitor models, we follow the time evolution to 200--500 ms after the bounce for the $9.6M_\odot$ and $15M_\odot$ models and up to the formation of a black hole for the $30M_\odot$ model. We follow the same procedure of simulations as described in \citet{2005ApJ...629..922S} and \citet{2007ApJ...667..382S}. We handle four neutrino species by assuming the same reactions for $\mu$-type and $\tau$-type neutrinos\footnote{See \citet{2017PhRvL.119x2702B} and \citet{2020PhRvD.102l3001F} for the treatment of muons.}. The weak reaction rates are based on the classic rates of \citet{1985ApJS...58..771B} with extensions of the nucleon--nucleon bremsstrahlung.  This setting is adopted partly to accommodate comparisons with the results of previous studies by \citet{2014PTEP.2014b3D05T}\footnote{See recent progress in neutrino reactions by \citet{2012ApJ...747...73L} and \citet{2018ApJ...853..170K} for example.}. For the $9.6M_\odot$ and $15M_\odot$ models, we adopt 1023 grids for the radial mass coordinate; these are larger than those in the previous simulations. This setting is helpful for examining explosion cases for the $9.6M_\odot$ model. The neutrino distribution function is discretized to 14 mesh points for the energy and six mesh points for the angle. For the $30M_\odot$ model, we adopt 255 grids for the radial mass coordinate using the rezoning method and we use 20 and four grids for neutrino energy and angle, respectively, which are the same as in the setting of \citet{2013ApJS..205....2N}.

\section{Supernova Models}
According to the standard scenario of core-collapse supernovae, an explosion is caused by the shock wave launched by the nuclear repulsion in the core of a massive star with the initial mass $\gtrsim\! 10M_\odot$ at the end of its life. Although the shock wave is considered to stall once, it eventually breaks through the stellar surface. For the shock revival, neutrinos will play an important role, as well as multidimensional effects such as the standing-accretion-shock instability and convection~\citep[for example]{2012PTEP.2012aA301K,2013RvMP...85..245B,2016ARNPS..66..341J}. Since the mass accretion rate of the outer layer affects the shock dynamics, the diversity in observed supernovae would stem from the variety of progenitors~\citep[for example]{2012ApJ...757...69U,2015PASJ...67..107N}.

\subsection{Collapse of $15M_\odot$ star}\label{sec:15m}
In this section, we adopt a $15M_\odot$ progenitor provided by \citet{1995ApJS..101..181W}, which is frequently employed as a benchmark for core-collapse simulation. \citet{2014PTEP.2014b3D05T} compared core-collapse results of this progenitor model between the Togashi EOS and Shen EOS, although the low-density region where the transition to the inhomogeneous phase occurs was matched to the Shen EOS. Furthermore, the core-collapse simulation of this model with the Shen EOS was described by \citet{2005ApJ...629..922S}. Following their numerical procedure, in this paper, we compute the collapse of the $15M_\odot$ progenitor model adopting the Togashi EOS, which has a low-density region including the nonuniform matter constructed consistently with the high-density region.

\begin{figure}[htbp]
\begin{center}
  \includegraphics[width=8cm]{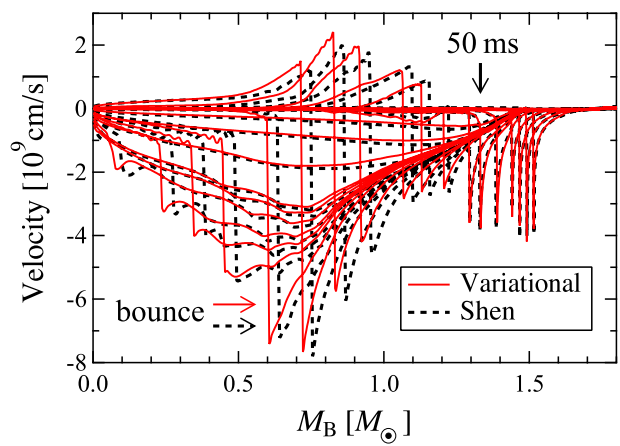}
\end{center}
  \caption{%
     Velocity profiles at selected times as a function of the baryon mass coordinate $M_\mathrm{B}$ for the models with $15M_\odot$. Red solid and black dashed lines are for the models with Togashi (variational) EOS and Shen EOS, respectively.
     }%
  \label{fig:15M_velocity}
\end{figure}

\begin{figure}[htbp]
\begin{center}
  \includegraphics[width=8cm]{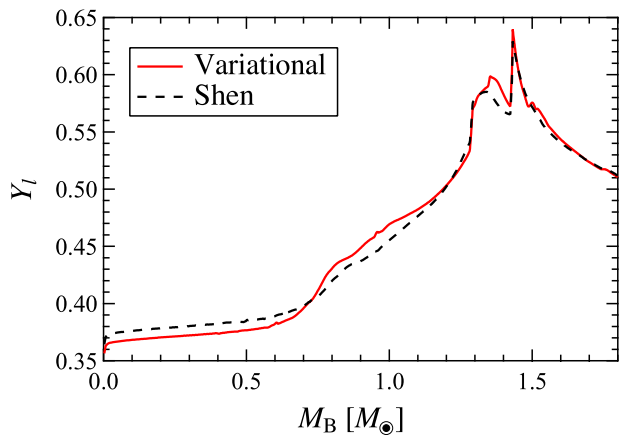}
\end{center}
  \caption{Profiles of the lepton fraction as a function of the baryon mass coordinate at the core bounce. The meanings of lines are the same as in Fig.~\ref{fig:15M_velocity}.%
     }%
  \label{fig:15M_bncYl}
\end{figure}

Figure \ref{fig:15M_velocity} shows the profiles of velocity in the overall dynamics from the gravitational collapse, the core bounce and the shock propagation. Both the Togashi and Shen EOS models show no explosion owing to the limitation of the spherical symmetry. The behavior of gravitational collapse of the Togashi EOS model is similar to that of the Shen EOS model because the free fall and the stalling of the shock wave similarly occur at late stages. However, the size of the inner core at around the core bounce is smaller for the Togashi EOS model. The size the of bounce core depends on the lepton fraction $Y_l$. In Fig.~\ref{fig:15M_bncYl}, we show $Y_l$ as a function of the baryon mass coordinate and observe that the Togashi EOS model has a smaller $Y_l$ than the Shen EOS model. Regarding the comparison between the Togashi and Shen EOS models, the difference in $Y_l$ accounts for the difference in the size of the inner core because the mass of a homologous core is proportional to $Y_l^2$ at the core bounce \citep{1983ApJ...270..735B}. 

\begin{figure}[htbp]
\begin{center}
  \includegraphics[width=8cm]{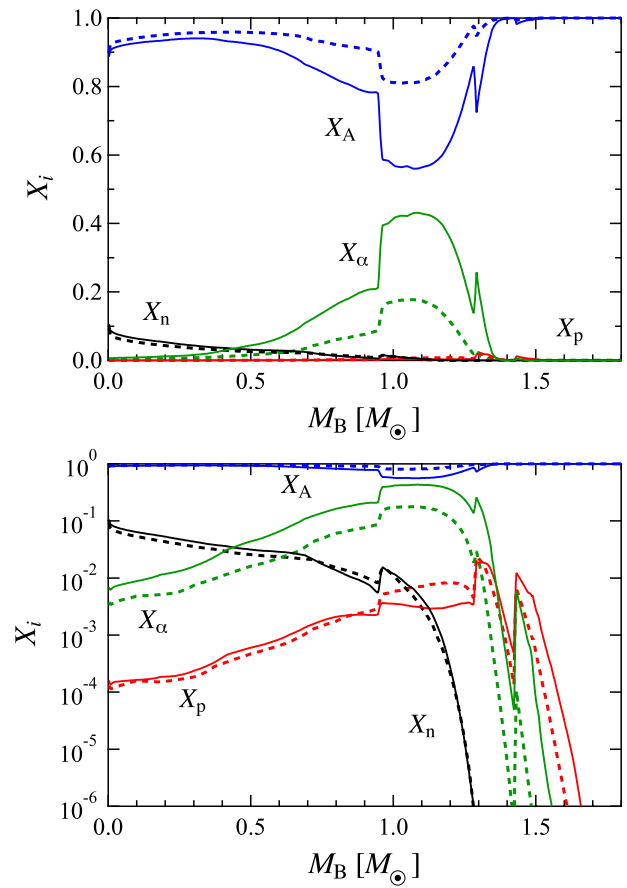}
\end{center}
  \caption{Mass fractions in the $15M_\odot$ models as a function of baryon mass coordinate at the time when the central density reaches $10^{12}~{\rm g~cm}^{-3}$. The lower plots are the same data as in the upper plots but shown in a log scale. Solid and dashed lines are for the models with the Togashi EOS and Shen EOS, respectively.%
     }%
  \label{fig:15M_preX}
\end{figure}

The lepton fraction is the sum of the electron and neutrino fractions. At the beginning of the collapse, $\nu_e$ emitted from electron capture can escape from the central core and the lepton fraction decreases. As the collapse proceeds, the matter density increases and neutrinos are gradually trapped. Eventually, neutrinos and electrons reach equilibrium and the lepton fraction stops evolving. The dominant opacity source for neutrinos is coherent scattering off heavy nuclei. In Fig.~\ref{fig:15M_preX}, we show the mass fraction at the time when the central density reaches $10^{12}~{\rm g~cm}^{-3}$. During the collapse stages, the mass fraction of nuclei in the Togashi EOS is smaller, which leads to a more efficient escape of neutrinos and a smaller amount of neutrino trapping. Thus, the Togashi EOS model has a smaller lepton fraction. Note that, in the case of proto-neutron star cooling \citep{2018PhRvC..97c5804N}, the Togashi EOS has a larger fraction of heavy nuclei than the Shen EOS, which is opposite to the case in Fig.~\ref{fig:15M_preX}. This is because the density is different in the two cases and, at low densities as in the prebounce stage, the dissociation of heavy nuclei in the Togashi EOS occurs at a temperature lower than that in the Shen EOS \citep{2017NuPhA.961...78T}.

\begin{figure}[htbp]
\begin{center}
  \includegraphics[width=8cm]{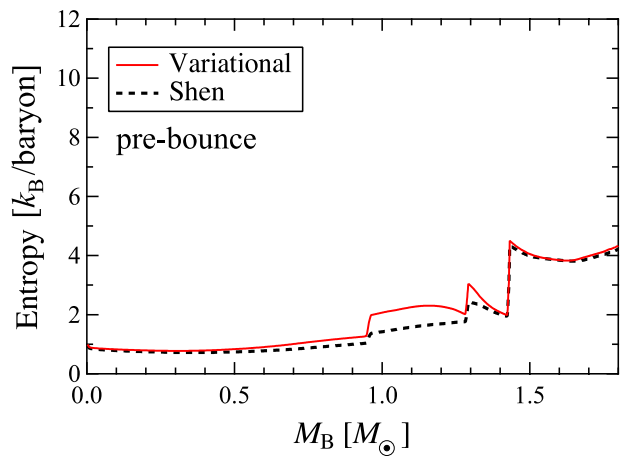}
\end{center}
  \caption{Entropy profiles as a function of the baryon mass coordinate at the time when the central density reaches $10^{12}~{\rm g~cm}^{-3}$. Red solid and black dashed lines are for the models with Togashi (variational) EOS and Shen EOS, respectively.%
     }%
  \label{fig:15M_preS}
\end{figure}

\begin{figure}[htbp]
\begin{center}
  \includegraphics[width=8cm]{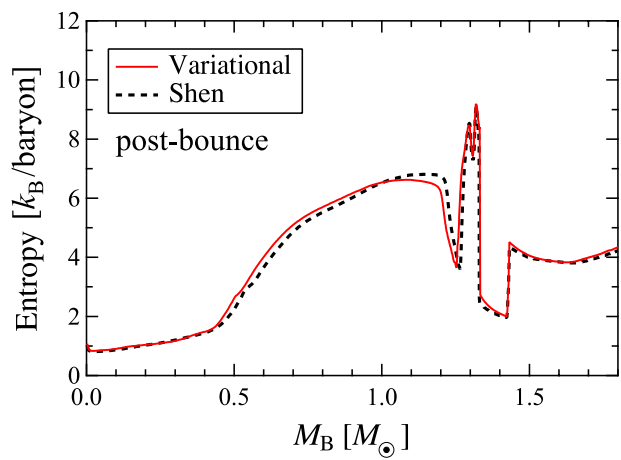}
\end{center}
  \caption{Same as Fig.~\ref{fig:15M_preS} except that the profiles at 50 ms after the core bounce are shown.%
     }%
  \label{fig:15M_postS}
\end{figure}

The difference in dissociation temperature for heavy nuclei is reflected also in the entropy. In Fig.~\ref{fig:15M_preS}, we show the entropy profile at the time when the central density reaches $10^{12}~{\rm g~cm}^{-3}$. We can see that the Togashi EOS model has a higher entropy than the Shen EOS model, especially in the range $1.0M_\odot \lesssim M_{\rm B} \lesssim 1.4M_\odot$, where $M_{\rm B}$ is the baryon mass coordinate. This difference stems from the construction of the initial conditions, which is performed so that the Togashi and the Shen EOS models have the same temperature profile. As shown in Fig.~15 in \citet{2017NuPhA.961...78T}, in the region where the mixing of alpha particles occurs, the Togashi EOS has a higher entropy than the Shen EOS if the temperature is fixed. Nevertheless, the shock wave resets the entropy. Figure \ref{fig:15M_postS} shows the profiles of entropy at $t_{\rm pb}=50$ ms, where $t_{\rm pb}$ is the time measured from the core bounce. The difference between the two models becomes small behind the shock wave, which is located at $M_{\rm B} \sim 1.3M_\odot$. Incidentally, according to \citet{2003PhRvL..91t1102H}, the behavior of the shock stall is unchanged even if the $Y_l$ profile is affected by improving the rate of electron capture on heavy nuclei. The evolution of the shock wave after $t_{\rm pb}=50$ ms is accordingly similar in both models, as seen in Fig.~\ref{fig:15M_velocity}.

\begin{figure*}[t]
\begin{center}
  \includegraphics[width=17cm]{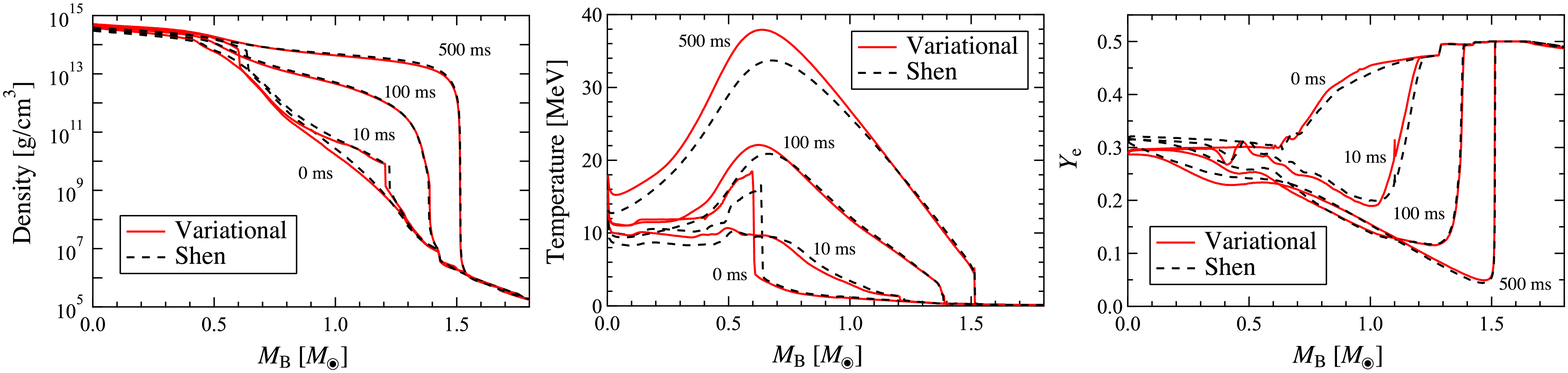}
\end{center}
  \caption{%
     Profiles of the density (left), temperature (center), and electron fraction (right) at selected times measured from the bounce as a function of the baryon mass coordinate for the models with $15M_\odot$. Red solid and black dashed lines are for the models with Togashi (variational) EOS and Shen EOS, respectively.
     }%
  \label{fig:15M_profiles}
\end{figure*}

We show in Fig.~\ref{fig:15M_profiles} the profiles of the density, temperature, and electron fraction at the core bounce and the post-bounce phase. Incidentally, the evolutions of the density and temperature profiles during the collapse are similar in the two EOS models. After the core bounce, the central density is higher for the Togashi EOS model because of the softness of the EOS, which is characterised by the compressible property. Accordingly, the temperature in the inner bounce core with the Togashi EOS is higher than that with the Shen EOS. These differences are attributed to the difference in the EOS at supra-nuclear densities, which is described in \citet{2014PTEP.2014b3D05T}.
\begin{figure*}[t]
\begin{center}
  \includegraphics[width=17cm]{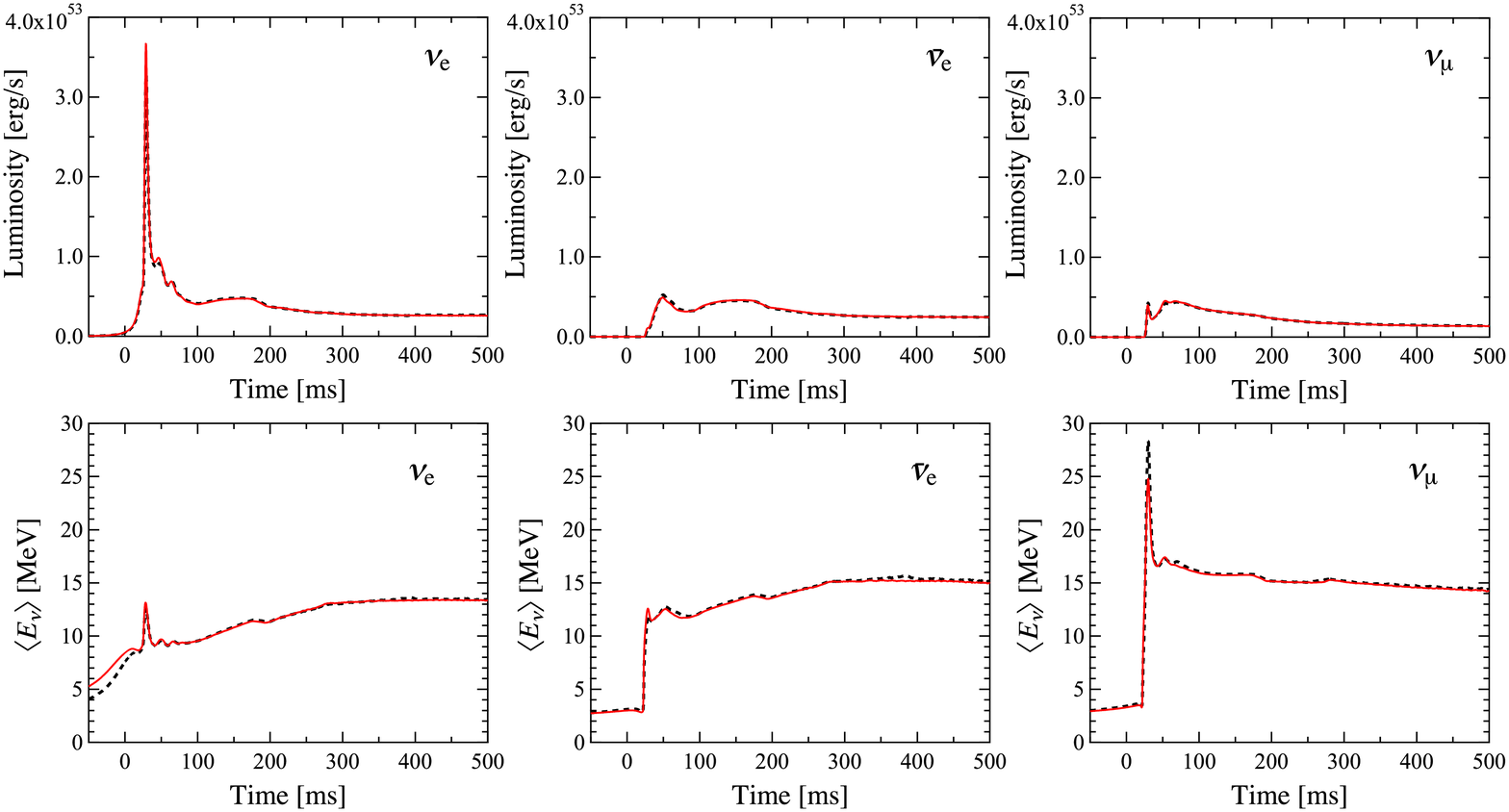}
\end{center}
  \caption{%
     Luminosity (upper panels) and average energy (lower panels) of emitted $\nu_e$ (left), $\bar\nu_e$ (center), and $\nu_\mu$ (right) for the models with $15M_\odot$ as a function of the time measured from the bounce. Red solid and black dashed lines are for the models with the Togashi EOS and Shen EOS, respectively.
     }%
  \label{fig:15M_lumiene}
\end{figure*}

Here, we focus on the profiles of the electron fraction $Y_e$ (right panel, Fig.~\ref{fig:15M_profiles}). At the core bounce, $Y_e$ in the central region of the Togashi EOS model is smaller than that in the Shen EOS model for the following two reasons: (1) $Y_l$ is smaller for the Togashi EOS model as already described, and (2) the Togashi EOS has a lower symmetry energy than the Shen EOS, which leads to a more neutron-rich system. On the contrary, $Y_e$ in the outer core region is larger for the Togashi EOS model. This slight difference arises in the collapse phase, during which the composition is different in the two EOS models; the fraction of free protons, which are the dominant source of electron capture, at around $M_{\rm B} \sim 1.0M_\odot$ in the Togashi EOS model is smaller than that in the Shen EOS model (lower plots, Fig.~\ref{fig:15M_preX}). Therefore, the electron capture is less efficient and more electrons remain for the Togashi EOS model. The difference in free proton fraction is again related to the mixing of alpha particles due to the dissociation of heavy nuclei. Nevertheless, the shock wave resets the composition. After the shock propagation, neutrinos in the core are trapped and their diffusion drives the evolution of the $Y_l$ profile. Accordingly, the post-shock evolution of $Y_e$ obeys the equilibrium condition.

We show in Fig.~\ref{fig:15M_lumiene} the luminosity and average energy of neutrino emission for three species ($\nu_e$, $\bar\nu_e$, and $\nu_\mu$) as a function of time. Note that the signals of $\nu_\mu$ and $\bar\nu_\mu$ are almost identical because they have the same reactions with a minor difference in coupling constants, and we assume that $\nu_\tau$ ($\bar\nu_\tau$) is the same as $\nu_\mu$ ($\bar\nu_\mu$) in our simulations. The neutrino emission is highly similar in the two EOS models because the evolution of profiles is similar, especially around the neutrino sphere in surface regions. This is because the central density is not high and the electron fraction is not low enough to cause any EOS difference. The difference in neutrino signal becomes more significant at later stages in the cooling of the proto-neutron star, as pointed out by \citet{2018PhRvC..97c5804N} and \citet{2019ApJ...887..110S}.  

\subsection{Collapse of $9.6M_\odot$ star}\label{sec:z9.6}
In this section, we adopt a nonrotating zero-metallicity $9.6M_\odot$ progenitor model of the iron core constructed by Heger\footnote{\url{http://2sn.org/firststars/znuc/presn/}}. Several authors have investigated the core collapse of this progenitor and reported its successful explosion even in spherically symmetric simulations. The Lattimer--Swesty EOS \citep{1991NuPhA.535..331L} with an incompressibility of 220~MeV (hereafter, LS220 EOS) was adopted by \citet{2015ApJ...801L..24M} and \citet {2017ApJ...850...43R}, while the DD2 EOS \citep{2010NuPhA.837..210H} was adopted by \citet{2020arXiv201016254M}. In this study, we employ the Togashi EOS and Shen EOS to compute the collapse of the $9.6M_\odot$ progenitor model.

\begin{figure}[htbp]
\begin{center}
  \includegraphics[width=6cm]{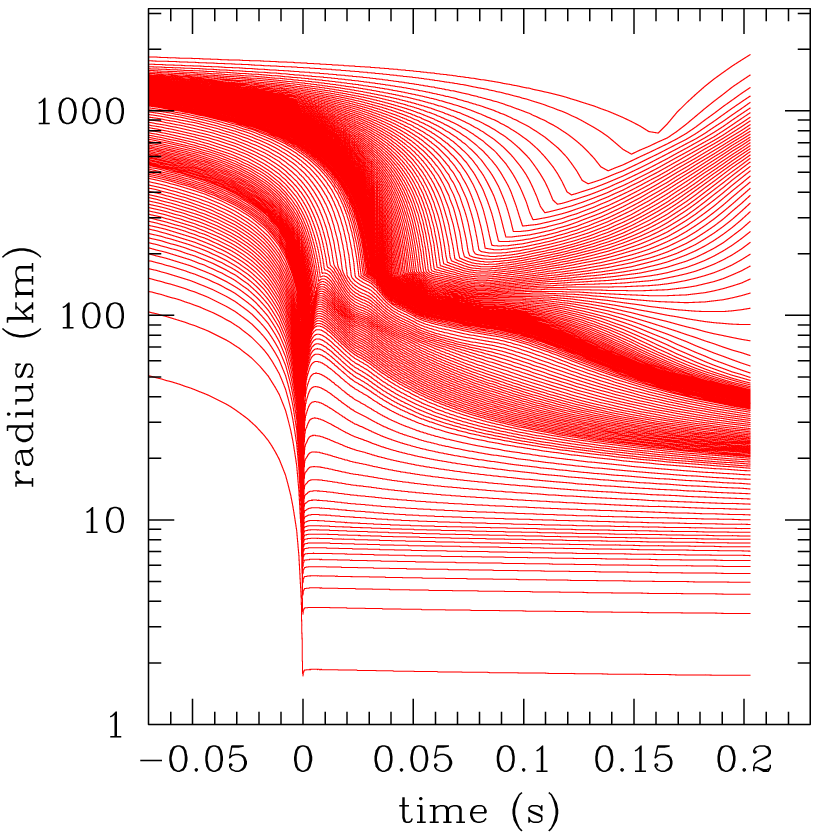}
\end{center}
  \caption{%
     Radial trajectories of mass elements as a function of time after the bounce in the progenitor model with $9.6M_\odot$ for the Togashi EOS. Lines are drawn for every seventh mass element while the width of mass mesh is nonuniform.
     }%
  \label{fig:z96gashi}
\end{figure}

\begin{figure}[htbp]
\begin{center}
  \includegraphics[width=6cm]{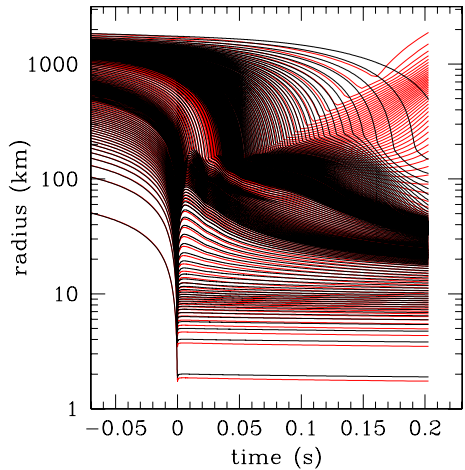}
\end{center}
  \caption{%
     Same as Fig.~\ref{fig:z96gashi} but black lines are for the Shen EOS model.
     }%
  \label{fig:z96shell}
\end{figure}

\begin{figure}[htpb]
\begin{center}
  \includegraphics[width=6cm]{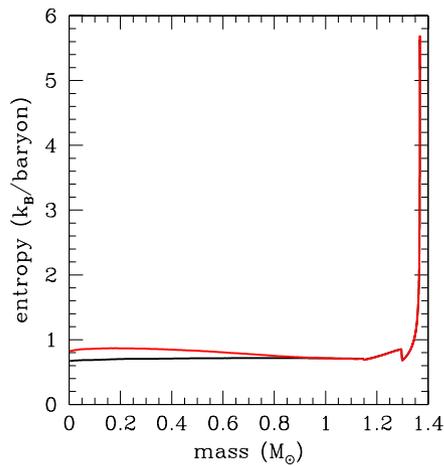}
\end{center}
  \caption{%
     Initial entropy profiles constructed using the Togashi EOS (red) and Shen EOS (black) for the progenitor model with $9.6M_\odot$.
     }%
  \label{fig:z96inits}
\end{figure}

\begin{figure}[htpb]
\begin{center}
  \includegraphics[width=6cm]{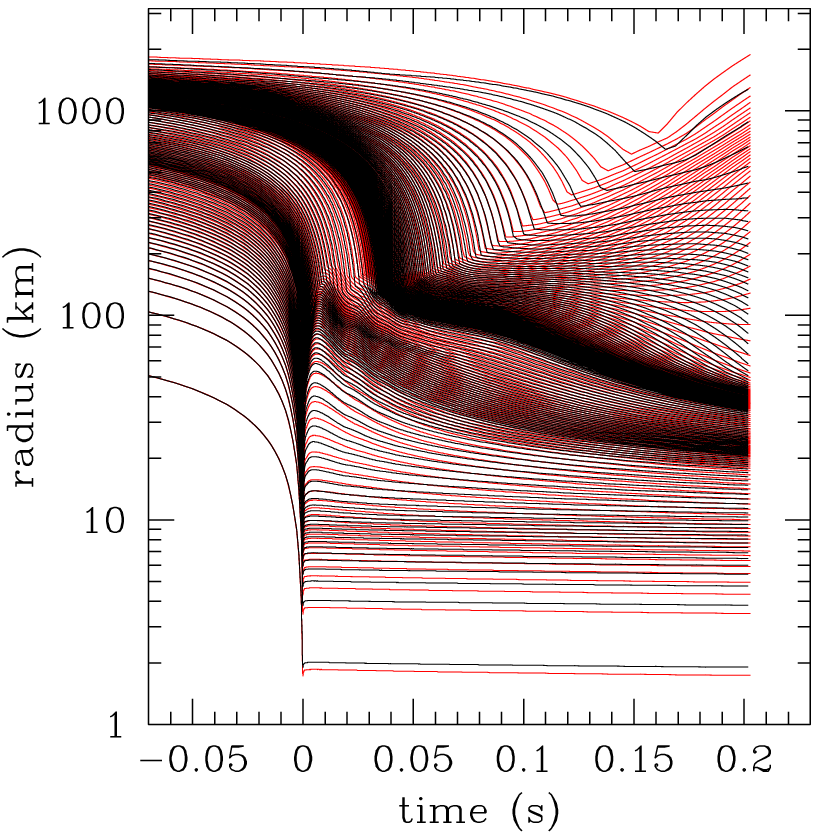}
\end{center}
  \caption{%
     Same as Fig.~\ref{fig:z96gashi} but black lines are for the model whose initial entropy profile was constructed using the Togashi EOS but core collapse was computed with the Shen EOS.
     }%
  \label{fig:z96shellv}
\end{figure}

In Fig.~\ref{fig:z96gashi}, we show the radial trajectories of mass elements for the core-collapse result obtained with the Togashi EOS. We can see that this model exhibits a successful explosion, as in other previous studies. In this model, the shock wave reaches a radius of 500~km at $t_{\rm pb}=137$~ms. In contrast, as shown in Fig.~\ref{fig:z96shell}, the core-collapse model with the Shen EOS does not lead to a successful explosion. In this model, the shock propagation is inhibited at a radius of 250~km and pushed back by the ram pressure of the infalling stellar envelope. Similarly to the case in \S\ref{sec:15m}, the models with the Togashi EOS and Shen EOS have different profiles of the initial entropy, as shown in Fig.~\ref{fig:z96inits}. According to \citet{2016MNRAS.460.2664S}, a small difference in initial entropy profile is crucial for a successful explosion\footnote{In \citet{2016MNRAS.460.2664S}, the progenitor models with a low central entropy were likely to explode, which is opposite to our case. Their factors for a successful explosion would be different from ours.}. In fact, if the initial entropy profile is set to be the same as the one constructed by the Togashi EOS, the collapse of the $9.6M_\odot$ model computed with the Shen EOS results in a successful explosion, as shown in Fig.~\ref{fig:z96shellv}. Incidentally, in this model, the shock wave reaches a radius of 500~km at $t_{\rm pb}=150$~ms.

\begin{figure}[htpb]
\begin{center}
  \includegraphics[width=6cm]{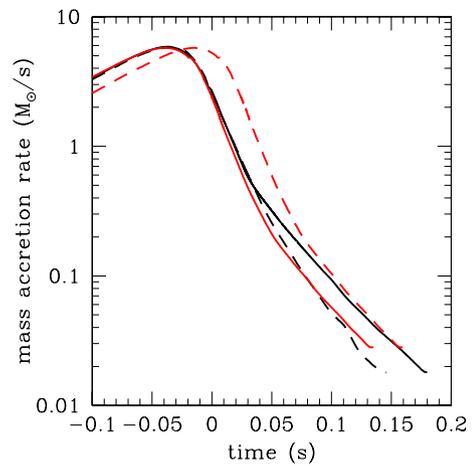}
\end{center}
  \caption{%
  Mass accretion rate at the radius of 500 km as a function of time after the bounce in the progenitor model with $9.6M_\odot$ for the Togashi EOS (red solid) and the Shen EOS (black solid). The black dashed line corresponds to the model whose initial entropy profile was constructed using the Togashi EOS but core collapse is computed with the Shen EOS. The red dashed line is the same as the red solid line but plotted by shifting forward by 24~ms, which is the bounce time difference between the Togashi and Shen EOS models.
     }%
  \label{fig:z96_mdot}
\end{figure}

Here, we seek to identify the is a key factor for a successful explosion in our models. In Fig.~\ref{fig:z96_mdot}, we show the mass accretion rate at the radius of 500~km. We can recognize that, for $t_{\rm pb}>30$~ms, the Togashi EOS model has a lower mass accretion rate than the Shen EOS model, which is advantageous to shock revival. Furthermore, the collapse using the Shen EOS having the initial entropy profile constructed with the Togashi EOS (black dashed line in Fig.~\ref{fig:z96_mdot}) results in a lower mass accretion rate than that using the initial entropy profile constructed with the Shen EOS. Therefore, in our models, the mass accretion rate determines whether or not the explosion is successful.

Now we focus on what determines the mass accretion rate. The difference in mass accretion rate stems from the difference in the time of the bounce measured from the onset of the simulation, which is 252~ms for the Togashi EOS model and 228~ms for the Shen EOS model. Although the difference in the pressure profile is minor, the Togashi EOS model has a higher initial entropy than the Shen EOS model in the inner region (Fig.~\ref{fig:z96inits}). Then, it takes a longer time for the core of the Togashi EOS model to collapse. On the other hand, the entropy in the outer region is the same for both models and the dynamics of the outer region is insensitive to the collapse of the inner core. Therefore, provided that the time is measured from the onset of the simulation, the infall time of the matter in the outer layer is similar for both models. This can be confirmed as follows. In Fig.~\ref{fig:z96_mdot}, the mass accretion rate for the Togashi EOS model is plotted by shifting toward back in time by 24~ms, which is the bounce time difference between the Togashi and Shen EOS models, along the red dashed line. This curve converges to the mass accretion rate for the Shen EOS model at a later time when the outer matter accretes onto the shock wave.

As noted in \S\ref{sec:15m}, the difference in the entropy profile is attributed to the difference in dissociation temperature for heavy nuclei between the Togashi EOS and Shen EOS models because the entropy per baryon increases due to the dissociation of heavy nuclei. Therefore, further investigation for the composition is important so as to prepare the initial condition of a core-collapse simulation. It would be preferred to start the simulation from the final stage of the stellar evolution and simultaneously solve the nuclear-burning network for following the composition as in \citet{2020arXiv201010506B}.

While we follow the evolution of the collapse of the progenitor model with $9.6M_\odot$ until $t_{\rm pb}=200$~ms, the outermost mass mesh of our Lagrangian code encounters the shock front. Hence, to continue the simulations for longer times, we should cover a further outer regime with lower densities than the coverage of the EOS tables, which is beyond the scope of this study. Thus, whether the collapse of the $9.6M_\odot$ model with the Shen EOS really results in a failure of the explosion is controversial in our simulation. The shock wave might revive afterward. Nevertheless, it is convincing that the Togashi EOS model in our simulation provides an example of the successful explosion of a low-mass progenitor. This result demonstrates that the Togashi EOS works well for explosive supernova simulations.

\begin{figure}[htpb]
\begin{center}
  \includegraphics[width=6cm]{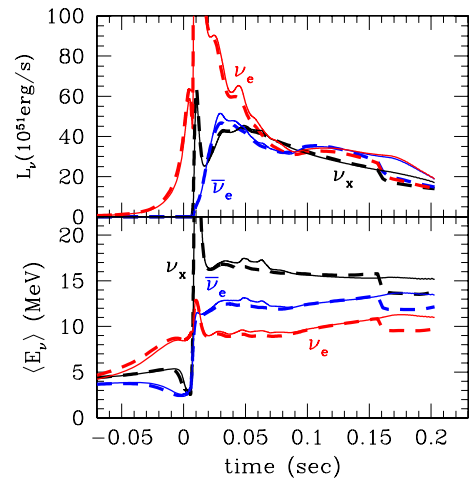}
\end{center}
  \caption{%
     Luminosities (upper plots) and average energies (lower plots) of the emitted $\nu_e$ (red), $\bar\nu_e$ (blue), and $\nu_x$ (black) as a function of time after the bounce for the models with $9.6M_\odot$, where $\nu_\mu$, $\bar\nu_\mu$, $\nu_\tau$, and $\bar\nu_\tau$ are collectively denoted as $\nu_x$. Thick dashed and thin solid lines are for the Togashi EOS and Shen EOS, respectively.
     }%
  \label{fig:z96el1}
\end{figure}

\begin{figure}[htpb]
\begin{center}
  \includegraphics[width=6cm]{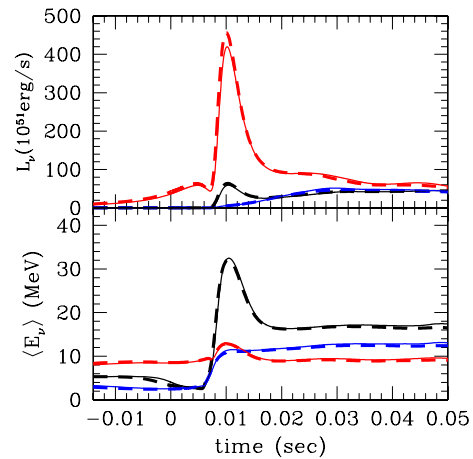}
\end{center}
  \caption{%
     Same as Fig.~\ref{fig:z96el1} but for around the time of the bounce.
     }%
  \label{fig:z96el2}
\end{figure}

In Figs.~\ref{fig:z96el1} and \ref{fig:z96el2}, we show the luminosities and average energies of neutrinos emitted from the $9.6M_\odot$ progenitor model. The sudden drop in the luminosity and average energy of the Togashi EOS model around $t_{\rm pb}=160$~ms corresponds to the time when the outermost mass mesh is blown off by the shock. In our code, the neutrino flux is evaluated by the distribution function of the outermost Lagrangian mesh in the comoving frame and the conversion to the observer frame is unimplemented. In ordinary cases dealt with using this code so far, the outermost mesh is located far enough to be static during the simulation time. By contrast, in the case of the $9.6M_\odot$ progenitor collapse, the outermost mesh moves with $O(0.1c)$, where $c$ is the speed of light, and the contribution of the Lorentz factor is not negligible. When the sign of the radial velocity is changed by the shock, artificial discontinuities appear in the plots for the neutrino luminosity and average energy. Nevertheless, we can confirm that the neutrino signals of the Togashi EOS model are similar to those of the Shen EOS model before the outermost mesh encounters the shock. Since the neutrino luminosity is the other key factor for shock revival besides the mass accretion rate \citep[for example]{1993ApJ...416L..75B,2005ApJ...623.1000Y}, the difference in mass accretion rate is again considered as essential for the successful explosion for our models.

\section{Black Hole Formation}
\label{sec:bhmodel}
If massive stars begin to collapse with high iron core mass, supernova explosions would fail and black holes form. Collapse without explosion is considered to be unlikely for stars with solar metallicity. However, for low-metallicity stars, mass loss in their evolutionary stages is inefficient. Thus, the very massive stars born in low-metallicity environments would completely collapse to black holes at their end. Furthermore, the resultant black hole mass is several tens of solar masses and consistent with the mass range of black holes observed via gravitational waves \citep{2016PhRvL.116f1102A,2018arXiv181112940T}.

In this section, we investigate black hole formation in the collapse of very massive stars. As is the case with ordinary supernovae, an enormous number of neutrinos are emitted from the black hole formation \citep[for example]{2006PhRvL..97i1101S}. We adopt the same progenitor model with the initial mass of $30M_\odot$ and the metallicity of $Z=0.004$ as used by \citet{2013ApJS..205....2N}. For this progenitor model, core-collapse simulations have already been performed using the Shen EOS \citep{2013ApJS..205....2N} and LS220 EOS \citep{2015ApJ...804...75N}.

As a result of the core-collapse simulation, we find that the time to black hole formation measured from the core bounce is 533~ms for the Togashi EOS, which is longer than that for the LS220 EOS (342~ms) but shorter than that for the Shen EOS (842~ms). In the case of black hole formation, a heavy envelope accretes onto the stalled shock and enlarges the central object. The central core begins to collapse again and becomes a black hole when its mass exceeds the critical value. This is analogous to the maximum mass of cold neutron stars. In our models, the recollapse is triggered when the baryonic mass of the central core becomes $2.49M_\odot$, $2.32M_\odot$, and $2.71M_\odot$ for the Togashi EOS, LS220 EOS, and Shen EOS, respectively. This result is consistent with those obtained by \citet{2020ApJ...894....4D}, where it is shown that, with increasing entropy, the maximum mass of hot neutron stars for the Shen EOS increases faster than that for the Togashi EOS, while the maximum mass of cold neutron stars for the Shen EOS is similar to that for the Togashi EOS.
\begin{figure*}[t]
\begin{center}
  \includegraphics[width=17cm]{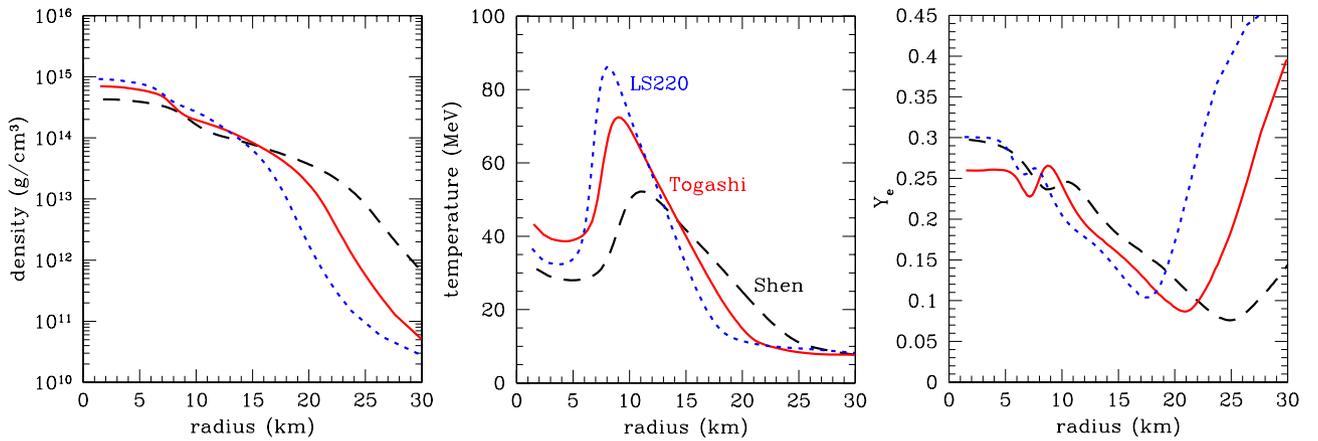}
\end{center}
  \caption{%
     Profiles of the density (left), temperature (center), and electron fraction (right) at 300~ms after the bounce for models with $30M_\odot$ and $Z=0.004$ in \citet{2013ApJS..205....2N}. Red solid, black dashed, and blue dotted lines represent the spectra for the models of the Togashi EOS, Shen EOS, and LS220 EOS, respectively.
     }%
  \label{fig:30M_snap}
\end{figure*}

In Fig.~\ref{fig:30M_snap}, we show the properties of the central core at $t_{\rm pb}=300$~ms. At that moment, the central core has a baryonic mass of $2.3M_\odot$ for all EOS models. We can recognize that the core radius\footnote{For instance, the core radius in the accretion phase is defined as the radius where the density is $10^{12}~{\rm g~cm}^{-3}$ in \citet{Schneider:2019shi}.} of the Togashi EOS model is larger than that of the LS220 EOS model but smaller than that of the Shen EOS model.
Incidentally, for cold neutron stars with typical masses, the Togashi EOS has a smaller radius than the LS220 EOS. The neutron star matter is less proton rich than the matter in the central core shown in Fig.~\ref{fig:30M_snap}. Therefore, the symmetry energy has a large impact on an EOS of neutron star matter. In particular, the slope parameter of symmetry energy, $L$, affects the neutron star radius: the EOS with a larger $L$ value has a larger radius of neutron stars provided that other properties are fixed \citep[for example]{2013ApJ...771...51L}. Note that the Togashi EOS has $L=39$~MeV and the LS220 EOS has $L=74$~MeV.

At $t_{\rm pb}=300$~ms, the central temperature of the Togashi EOS model is higher than that of the LS220 EOS model while the central density is the opposite (Fig.~\ref{fig:30M_snap}). It can be interpreted as follows. As discussed in \citet{Schneider:2019shi}, the thermal pressure of baryons obeys $P^{\rm th}_b \propto M^\ast T^2$, where $M^\ast$ is the effective mass of nucleons and $T$ is the temperature. While $M^\ast$ is assumed to be equal to the rest mass in the LS220 EOS, the density and temperature dependences of $M^\ast$ are considered and $M^\ast$ is lower than the rest mass in the Togashi EOS. Therefore, for supplying the thermal pressure, the temperature tends to become higher for the Togashi EOS model than that in the case with the assumption that $M^\ast$ is equal to the rest mass.

\begin{figure}[htbp]
\begin{center}
  \includegraphics[width=6cm]{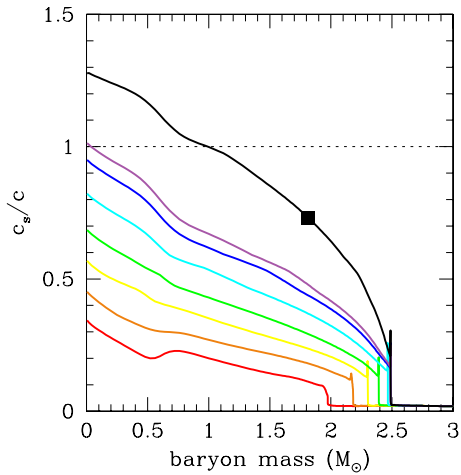}
\end{center}
  \caption{%
     Profiles of the sound velocity in unit of the speed of light as a function of baryon mass coordinate for the Togashi EOS model with $30M_\odot$ and $Z=0.004$ in \citet{2013ApJS..205....2N}. The lines correspond, from bottom to top, to $t_{\rm pb}= 100~{\rm ms}$, $200~{\rm ms}$, $300~{\rm ms}$, $400~{\rm ms}$, $500~{\rm ms}$, $532.56~{\rm ms}$ (onset of second collapse), $533.12~{\rm ms}$ (causality violation), and $533.36~{\rm ms}$ (apparent horizon formation), where $t_{\rm pb}$ is the time measured from the bounce. The square shows the location of the apparent horizon.
     }%
  \label{fig:30M_cs}
\end{figure}

In this study, we compute fully general relativistic hydrodynamics under spherical symmetry until the apparent horizon formation\footnote{In our simulation code \citep{1997ApJ...475..720Y}, we adopt the Misner--Sharp metric \citep{1964PhRv..136..571M}, in which the condition $2G\tilde{m}/rc^2>1$ is satisfied in the trapped region with the circumference radius $r$ and the gravitational mass $\tilde{m}$, and this metric enable us to follow the dynamics until the apparent horizon formation as in \citet{1966PhRv..141.1232M}.}, which is the sufficient condition for the existence of an event horizon. Once the gravitational instability is again triggered by the general relativity, the central core collapses to a black hole within the dynamical timescale: $1/\sqrt{G\rho} \sim O(0.1)$~ms, where $G$ is the gravitational constant and $\rho$ is the density. During the second collapse, the central density increases suddenly. At supranuclear densities, the Togashi EOS violates the causality owing to its nonrelativistic nature. In particular, for the cold neutron star EOS, the critical density for causality violation is lower than the central density of maximum-mass neutron stars \citep{2007NuPhA.791..232K,2013NuPhA.902...53T}. In Fig.~\ref{fig:30M_cs}, we show the profiles of the sound velocity for our model of black hole formation with the Togashi EOS. Here, we can see that the sound velocity exceeds the speed of light after the onset of the second collapse but before the apparent horizon formation. Therefore, we presume that the evolutions of the central core with matter accretion and its neutrino emission are reasonably computed for the quasi-static phase, although the dynamics leading to the black hole formation is unconvincing. Furthermore, the apparent horizon is formed outside the region where the causality is violated. Incidentally, the central density at the moment of causality violation is $1.59 \times 10^{15}~{\rm g~cm}^{-3}$, whereas the critical density is $1.46 \times 10^{15}~{\rm g~cm}^{-3}$ for the cold neutron star EOS. Actually, at such high densities, possible phase transitions such as quark deconfinement would soften the EOS and hasten the apparent horizon formation \citep[for example]{2008PhRvD..77j3006N}.

\begin{figure*}[t]
\begin{center}
  \includegraphics[width=17cm]{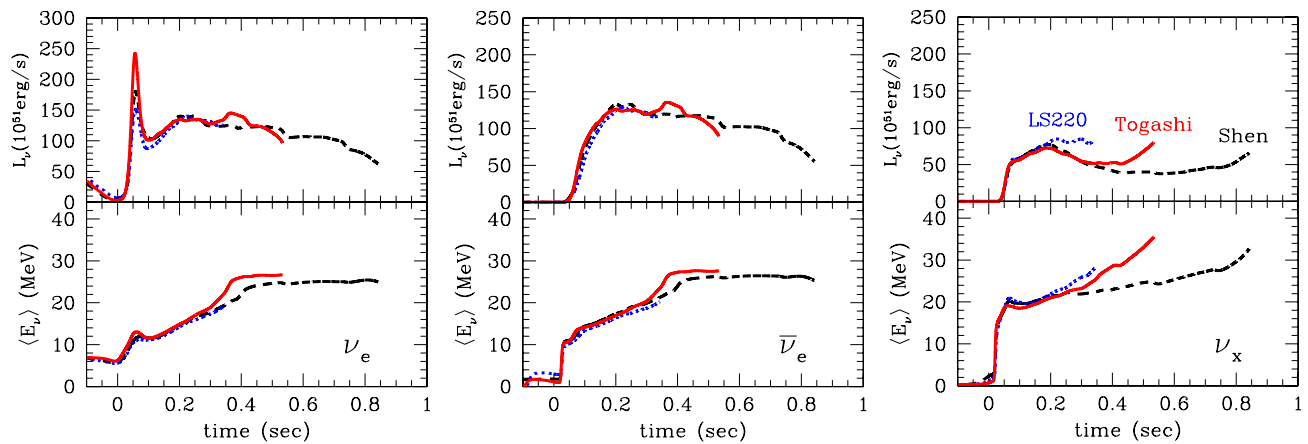}
\end{center}
  \caption{%
     Luminosities and average energies of $\nu_e$ (left), $\bar\nu_e$ (center), and $\nu_x$ (right) for models with $30M_\odot$ and $Z=0.004$ in \citet{2013ApJS..205....2N}, where $\nu_\mu$, $\bar\nu_\mu$, $\nu_\tau$, and $\bar\nu_\tau$ are collectively denoted as $\nu_x$. Red solid, black dashed, and blue dotted lines represent the spectra of the Togashi EOS, Shen EOS, and LS220 EOS, respectively.
     }%
  \label{fig:30M_enelumi}
\end{figure*}

The central core maintains the neutrino emission until a black hole is formed. The luminosities and average energies of neutrinos emitted from the black hole formation are shown in Fig.~\ref{fig:30M_enelumi} as a function of time after the bounce. The endpoints of the curves correspond to the times of black hole formation and we can regard the sharp truncation of the neutrino flux as an observational signature of black hole formation \citep{2001PhRvD..63g3011B,Li:2020ujl}. Comparing the different EOS models, we can see that the neutrino signals are similar to each other up to the time of black hole formation, especially for $\nu_e$ and $\bar\nu_e$. Since the energy of emitted neutrinos originates from the gravitational potential released by accreted matter, the neutrino luminosity is principally determined by the mass accretion rate onto the stalled shock, which resides in the region with subnuclear density. As for $\nu_x$, the average energy is higher for more compressible EOS models because the $\nu_x$ signal is sensitive to the inner region because of its long mean free path. Nevertheless, the difference in EOS, which mainly affects the central core with the supranuclear density, appears mainly in the duration of neutrino emission and its effect on the neutrino signal is not significant up to the black hole formation. We also note that, compared to the cases for the $15M_\odot$ and $9.6M_\odot$ models, the peaks in the profile of the neutrino luminosity and mean energy are rounded. This is because the number of mesh points in the radial mass coordinate and neutrino distribution function is different and the present black-hole-formation model has lower resolutions for the spatial and neutrino-angle grids (\S\ref{sec:sumiyoshicode}). In addition, since this model has the larger radius of the outer boundary, where the neutrino flux is evaluated, the luminosity peak due to the neutronization is delayed compared to other models.

\begin{figure}[htbp]
\begin{center}
  \includegraphics[width=6cm]{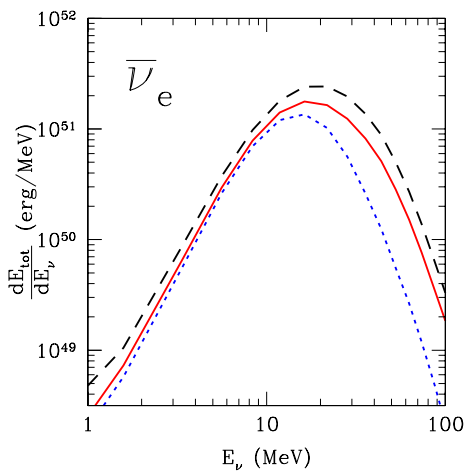}
\end{center}
  \caption{%
     Time-integrated spectra of $\bar\nu_e$ for models with $30M_\odot$ and $Z=0.004$ in \citet{2013ApJS..205....2N}. Red solid, black dashed, and blue dotted lines represent the spectra of the Togashi EOS, Shen EOS, and LS220 EOS, respectively.
     }%
  \label{fig:30M_spec}
\end{figure}

The time-integrated spectra of $\bar\nu_e$ are shown in Fig.~\ref{fig:30M_spec} and the total and mean energies of emitted neutrinos are summarized in Table~\ref{table:30M_neutrino}. As seen, the total energy of emitted neutrinos is larger for the EOS model with the longer interval between the bounce and the black hole formation, which corresponds to the duration of neutrino emission. Furthermore, the EOS model with the longer duration of neutrino emission has a high mean neutrino energy. This is because heating by accretion shock continues until the black hole formation and, thereby, the mean neutrino energy increases in the later phase. These differences would affect the flux and spectrum of supernova relic neutrinos \citep[for example]{2015ApJ...804...75N}. Incidentally, the numerical data of emitted neutrinos for the Togashi EOS model are publicly available on the Web\footnote{\url{http://asphwww.ph.noda.tus.ac.jp/snn/}} along with those for the other EOS models \citep{2013ApJS..205....2N}.

\begin{table}[htbp]
\tbl{Properties of emitted neutrinos for models with $30M_\odot$ and $Z=0.004$ in \citet{2013ApJS..205....2N}. 
\footnotemark[$*$] }{%
\begin{tabular}{@{}lcccccc@{}}  
\hline\noalign{\vskip3pt} 
 & $\langle E_{\nu_e} \rangle$ & $\langle E_{\bar \nu_e} \rangle$ & $\langle E_{\nu_x} \rangle$ & $E_{\nu_e, {\rm tot}}$ & $E_{\bar \nu_e, {\rm tot}}$ & $E_{\nu_x, {\rm tot}}$ \\
EOS & \multicolumn{3}{c}{(MeV)} & \multicolumn{3}{c}{($10^{52}$ erg)} \\  [2pt] 
\hline\noalign{\vskip3pt} 
   LS220 & 12.5 & 16.4 & 22.3 & 4.03 & 2.87 & 2.11 \\
   Togashi & 16.1 & 20.4 & 23.4 & 6.85 & 5.33 & 2.89 \\
   Shen & 17.5 & 21.7 & 23.4 & 9.49 & 8.10 & 4.00 \\  [2pt] 
\hline\noalign{\vskip3pt} 
\end{tabular}}\label{table:30M_neutrino}
\begin{tabnote}
\hangindent6pt\noindent
\hbox to6pt{\footnotemark[$*$]\hss}\unskip%
 With the total energy $E_{\nu_i, {\rm tot}}$ and total number $N_{\nu_i, {\rm tot}}$ of the emitted $\nu_i$ until the black hole formation, the mean neutrino energy is defined as $\langle E_{\nu_i} \rangle \equiv E_{\nu_i, {\rm tot}}/N_{\nu_i, {\rm tot}}$ where $\nu_x=\nu_\mu=\bar \nu_\mu=\nu_\tau=\bar \nu_\tau$. 
\end{tabnote}
\end{table}

\section{Concluding remarks}
\label{sec:concl}
In this study, we performed core-collapse simulations using the Togashi EOS, which is constructed on the basis of the variational many-body theory, and obtained the resultant neutrino signal. For this purpose, we investigated three cases: the $15M_\odot$ progenitor for a benchmark supernova model, the $9.6M_\odot$ progenitor for a low-mass supernova model with a successful explosion, and the $30M_\odot$ progenitor for a black hole formation. The core-collapse results were compared with those obtained with other EOS models.

For supernova simulations, the impact of the EOS is usually not significant either to the dynamics or the neutrino signal on the time scale of $O(100)$~ms. This is because the density is not high and the electron fraction is not low in the central core, as already described in the previous work \citep{2014PTEP.2014b3D05T}. Note that the publicly available models of the nuclear EOS are constructed to be consistent with the constraints from terrestrial nuclei. Nevertheless, the EOS difference in the low-density region including heavy nuclei affects the composition and size of the inner core at around the bounce. While this variation would be usually reset by the shock wave, we have found that it can play an important role in the fate of the shock wave propagation for the low-mass progenitor model by altering the behavior of the outer layer relative to the inner core.

For the simulations of black hole formation, the EOS determines the interval time from the bounce to the black hole formation because the maximum mass of hot neutron stars depends on the EOS. This is reflected in the duration of the neutrino emission. Incidentally, the Togashi EOS violates causality in the high-density region owing to its nonrelativistic formalism. In our numerical simulation, the causality violation occurs after the onset of the second collapse to the black hole formation and its region is immediately enveloped by the apparent horizon.

For both the low- and high-mass progenitor models, we confirmed that the Togashi EOS works well in the core-collapse simulations including neutrino emission. Since our simulations were carried out assuming spherical symmetry, it is preferable that the same models be applied to multidimensional simulations\footnote{See, for example,  \citet{2019ApJ...880L..28N}, \citet{2020ApJ...902..150H}, and \citet{2020ApJ...903...82I} regarding applications of the EOS based on variational many-body theory and EOS effects.}. The neutrino reaction rates will be modified particularly for the nucleon bremsstrahlung to contain the consistent correlation function with the Togashi EOS in future work. We believe that these issues deserve further investigation.

\begin{ack}
  The authors are grateful to Yudai Suwa and Hideyuki Suzuki for insightful suggestions and encouragements. In this work, numerical computations were partly performed on the supercomputers at Research Center for Nuclear Physics in Osaka University. Yukawa Institute of Theoretical Physics in Kyoto University, Nagoya University, the University of Tokyo, JLDG on SINET of NII, and Computing Research Center in KEK (the Particle, Nuclear and Astro Physics Simulation Program; Nos. 2019-002, 2020-004) are also acknowledged for providing the high performance computing resources. This work was partially supported by Grants-in-Aid for Scientific Research (JP18K13551, JP19K03837, JP20H01905, JP20K03973) and Grants-in-Aid for Scientific Research on Innovative Areas (JP17H06357, JP17H06365, JP19H05802, JP19H05811) from the Ministry of Education, Culture, Sports, Science and Technology (MEXT), Japan.
\end{ack}

\bibliography{bib.bib}
\end{document}